\documentclass[aps,prb]{revtex4}
\usepackage{graphicx}

\begin{document}
\pagestyle{empty}
\title{Quantum Vavilov-Cherenkov radiation from shearing  two transparent dielectric plates}

\author{A.I.Volokitin$^{1,2,3}$\footnote{Corresponding author.
\textit{E-mail address}:alevolokitin@yandex.ru}    and B.N.J.Persson$^1$}
 \affiliation{$^1$Peter Gr\"unberg Institut,
Forschungszentrum J\"ulich, D-52425, Germany}

\affiliation{
$^2$Samara State Technical University, Physical Department, 443100 Samara, Russia}
\affiliation{
$^3$Samara State Aerospace University, Physical Department, 443086 Samara, Russia}

\begin{abstract}
Using a fully relativistic theory we study the quantum Vavilov-Cherenkov radiation  and quantum friction occurring during relative 
sliding of the two   transparent dielectric plates with the refractive index  $n$.   These phenomena occur above  
the threshold velocity $v_c=2nc/(n^2+1)$.   Close to the  threshold velocity they are dominated by the contribution from $s$--polarized electromagnetic waves, which agrees with the approximate 
(relativistic) theory by Pendry (J. Mod. Opt. \textbf{45}, 2389 (1998)).  
However, in the ultra relativistic case ($v\rightarrow c$), the contributions from both polarisations are strongly enhanced in the comparison with the approximate theory, and 
a new
contribution occurs from the  mixing of the electromagnetic waves with the different polarization.   The numerical results are supplemented by an analytical treatment close to the threshold velocity and the light velocity.
\end{abstract}

\maketitle


\vskip 5mm

\section{Introduction}
Quantum  fluctuations of the electromagnetic field manifest themself in a wide variety of fields of physics. 
For example, the Lamb shift of atomic spectrum and anomalous magnetic moment of the electron were explained with the help of this idea.  
Most directly, these fluctuations are manifested through the Casimir effect. In the late 1940s Hendrik Casimir predicted \cite{Casimir1948} that two macroscopic non-magnetic bodies 
with no net electric charge (or charge moments) can experience an attractive force much stronger than gravity. The existence of this force is one of the direct 
macroscopic manifestations of quantum mechanics.
Hendrik Casimir based his prediction on a simplified model involving two parallel perfectly conducting plates separated by vacuum. 
A unified theory of both the van der Waals and Casimir forces between plane parallel material plates, in thermal equilibrium and separated by a vacuum gap, was developed by Lifshitz (1955) 
\cite{Lifshitz1955}. To calculate the interaction force Lifshitz used Rytov’s theory of the fluctuating electromagnetic field \cite{Rytov1989}.  
At present the interest of Casimir forces is increasing because they 
dominate the interaction between nanostructures, and are offen responsible for the stiction  between moving parts in small devices such as micro- and nanoelectromechanical systems, and can be 
considered as practical mechanisms for the actuation of such devices.  Due to this practical interest, and the fast progress in force detection techniques, experimental and theoretical
 investigations of Casimir forces have experienced an extraordinary “renaissance” in the past few years (see \cite{Dalvit2011} for a review and the references therein).

Another manifestation of  quantum fluctuations of the electromagnetic field is the non-contact quantum friction between bodies in 
relative motion. Friction 
is usually a very complicated process. The simplest case consists of two flat surfaces, separated by a vacuum gap,  sliding relative to each 
other at $T=0$ K, where the friction is generated by the relative movement of quantum  
fluctuations \cite{PendryJPCM1997,PendryJMO1998,VolokitinPRL2003,VolokitinPRB2003,VolokitinJPCM1999,VolokitinRMP2007,VolokitinPRB2008,VolokitinPRL2011}. The thermal 
and quantum fluctuation of the current density in one body induces a current density 
in other body; the interaction between these current densities is the origin of the 
Casimir interaction. When two bodies are in relative motion, the induced current will 
lag slightly behind the fluctuating current inducing it, and this is the origin of the 
Casimir friction. At present the Casimir friction, with its limiting case - quantum friction, are 
actively discussed as one of the possible mechanisms of noncontact friction between bodies in absence of direct 
mechanical contact between them \cite{VolokitinRMP2007,Fundamentals2015}. The Casimir friction  was studied in the configurations plate-plate \cite{PendryJPCM1997,PendryJMO1998,VolokitinJPCM1999,VolokitinPRL2003,VolokitinPRB2003,VolokitinRMP2007,VolokitinPRB2008,VolokitinPRL2011},  neutral particle-plate \cite{VolokitinRMP2007,TomassonePRB1997,VolokitinPRB2002,DedkovPLA2005,DedkovJPCM2008,BartonNJP2010,BrevikEntropy2013,BrevikEPJD2014,KardarPRD2013,
DalvitPRA2014,VolokitinNJP2014,HenkelNJP2013,HenkelJPCM2015},  and neutral particle-blackbody radiation  \cite{VolokitinRMP2007,VolokitinPRB2008,HenkelNJP2013,MkrtchianPRL2003,DedkovNIMPR2010,
JentschuraPRL2012,JentschuraPRL2015,VolokitinPRA2015}. While the predictions of the theory for the Casimir forces were verified in many experiments \cite{Dalvit2011}, the detection of the 
Casimir friction  is still challenging problem for  experimentalists. However, the frictional  drag between quantum wells \cite{GramilaPRL1991,SivanPRL1992} and  graphene sheets 
\cite{KimPRB2011,GeimNaturePhys2012}, and the current-voltage dependence of nonsuspended graphene on the surface of the polar dielectric SiO$_2$ \cite{FreitagNanoLett2009}, were accurately described using the 
theory of the Casimir friction \cite{VolokitinPRL2011,VolokitinJPCM2001b,VolokitinEPL2013}. At present the frictional drag experiments \cite{GramilaPRL1991,SivanPRL1992,KimPRB2011,GeimNaturePhys2012,FreitagNanoLett2009} were performed for weak electric field when the induced drift motion of the free carriers is smaller than 
the threshold velocity for quantum friction. Thus in these experiments the frictional drag is dominated by the contributions from  thermal fluctuations. However, the measurements of the current-
voltage dependence \cite{FreitagNanoLett2009} were performed for high electric field, where the drift velocity is above the threshold velocity, and where the frictional drag is dominated by  quantum fluctuations \cite{VolokitinPRL2011,VolokitinEPL2013}. 

Quantum friction is associated with creation of  excitations of  different kind. 
For transparent dielectrics such excitations are photons and quantum friction 
is associated with quantum Vavilov-Cherenkov radiation.  Thus there is a close connection between quantum 
friction and the quantum Vavilov-Cherenkov 
radiation -- both of these phenomena are related to the  anomalous Doppler effect 
\cite{Frank1943,Ginzburg1945, Ginzburg1996,PendryJPCM1997,VolokitinRMP2007,VolokitinPRL2011,KardarPRA2013}. 
Quantum  Vavilov-Cherenkov radiation was first  described by  Frank \cite{Frank1943} and   Ginzburg and  Frank \cite{Ginzburg1945} (see also \cite{Tamm1959,Ginzburg1996} for 
review of these work). If an object has no internal degrees of freedom (e.g., a point charge), then the energy of the radiation is determined by the change of 
the  kinetic energy of the object. However, if an object has  internal degrees of freedom (say, an atom), then two types  of radiation 
 are possible.  If the frequency of the radiation in the \textit{lab} reference frame $\omega >0$, then in  the rest frame of an object, due to the Doppler effect, 
the frequency of the radiation $\omega^{\prime}=\gamma(\omega-k_xv)$, where $\gamma = \sqrt{1-\beta^2}$, $\beta=v/c$. In the normal Doppler effect region, 
when $\omega^{\prime}>0$, the radiated energy is determined by the decrease of the internal energy. For example, for an atom the state may changes from the excited state $|1>$ to the 
ground state 
$|0>$. The region of the \textit{anomalous} Doppler effect corresponds to $\omega^{\prime}<0$ in which case an object becomes excited when it radiates. For example, an atom could  
 experience the transition from the ground state $|0>$ to the excited state $|1>$ when it radiates. In such a case energy conservation requires that the energy  of the radiation and of the excitation 
result from a
 decrease of the kinetic energy of the object. That is, the self-excitation of a system is accompanied by a slowing down of the motion of the object as a whole. 
For a neutral object the interaction of the object 
with the matter is determined by the fluctuating electromagnetic field due to the quantum fluctuations inside the object. 

 While a constant translational motion  requires for the emission of the radiation at least two bodies in relative motion (otherwise it is not possible due to Lorenz invariance), a single accelerated object 
can radiate and experience friction. Quantum fluctuations of the electromagnetic field are determined by virtual photons that are continuously created and annihilated in the vacuum. Using a metal mirror in   
accelerated motion, with velocities  near the light velocity, virtual photons can be converted into real photons, leading to radiation emitted by the mirror. This is the dynamic Casimir effect \cite{Moore1970,Danies1976,Dalvit2011}; which recently 
 was observed in a superconducting waveguide \cite{Wilson2011}. Radiation can be also emitted by a rotating object \cite{ManjavacasPRL2010,ManjavacasPRA2010,KardarPRL2012,KardarPRA2014,BercegolPRL2015}. In fact this phenomenon is closely related to the  prediction 
by Zel'dovich \cite{ZeldovichPJETP1971} of  an amplification of  
certain waves during scattering from a rotating body. Rotational quantum friction is strongly enhanced close to a dielectric substrate \cite{PendryPRL2012}

Pendry was the first to studied quantum friction in detail \cite{PendryJPCM1997} in the non-relativistic and non-retarded limit. In Ref. \cite{PendryJMO1998}  Pendry   applied 
a new   formalism developed for quantum friction 
to estimate the  emission of light
occurring during relative 
sliding of  two   transparent dielectric plates. To take into account some relativistic effects  the reflection amplitudes for the moving surface was 
taken as   the reflection amplitudes in the rest reference frame for this surface.  The relation between the frequency and the wave-vector, which determine the arguments of 
these reflection amplitudes were  determined  in the different reference 
frame  by the Lorenz transformation. In Ref. \cite{VolokitinPRB2008} we developed a fully relativistic theory of quantum friction from which follows that the theory 
by Pendry \cite{PendryJMO1998} is accurate only  to  order $(v/c)^2$. Using a toy model the link between quantum Vavilov-Cherenkov radiation and quantum friction was also discussed 
within framework of  a non-relativistic theory  in Ref. \cite{KardarPRA2013}. 

In the present article  we use a fully relativistic theory  \cite{VolokitinPRB2008} to study 
quantum Vavilov-Cherenkov and quantum friction during shearing of two transparent dielectric plates. 
Numerical calculations are supplemented  by an analytical study close to the threshold $v_c=2nc/(n^2+1)$ and light velocities. Close to the threshold 
velocity $v_c$ our results agree with the results obtained by Pendry.  In the ultra relativistic case ($v\rightarrow c$) there is a dramatic enhancement of the contributions to quantum friction  
related to radiation of both polarisations, and a new contribution occurs from  polarization mixing.

\begin{figure}[tbp]
\includegraphics[width=0.40\textwidth]{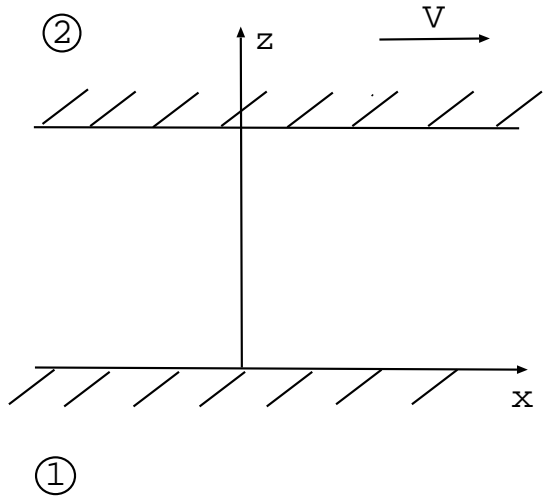}
\caption{Two semi-infinite bodies with plane parallel surfaces separated by
a distance $d$. The upper solids moves parallel to other with the velocity $v$. }
\label{Fig1}
\end{figure}

\section{Photon emission and Anomalous Doppler effect}
We consider two semi-infinite solids having flat parallel surfaces separated
by a distance $d$ and moving with the velocity $v$ relative to each other,
see Fig. \ref{Fig1}.
We introduce the two reference frames $K$ and $K^{\prime }$ with
coordinate axes $xyz$ and $x^{\prime }y^{\prime }z^{\prime }$. In the $K$
frame body \textbf{1} is at rest while body \textbf{2} is moving with the
velocity $v$ along the $x-$ axis (the $xy$ and $x^{\prime }y^{\prime }$ planes
are in the surface of body \textbf{1, }$x$ and $x^{\prime}$- axes have the
same direction, and the $z$ and $z^{\prime}-$ axes point toward body \textbf{%
2}). In the $K^{\prime}$ frame body \textbf{2} is at rest while body
\textbf{1} is moving with velocity $-v$ along the $x-$ axis.
If in  body \textbf{2}, in the  $K^{\prime }$ frame,   excitation occurs with frequency 
$\omega_{\alpha 2}^{\prime}(q^{\prime})$ and wave vector $\mathbf{-q}^{\prime}=
(-q_x^{\prime},-q_y,0)$, then in the laboratory reference frame $K$ this excitation will 
have the frequency  $\omega_{\alpha 2}(-q_x,-q_y) = \gamma(\omega_{\alpha 2}^{\prime}
(q^{\prime})-q_x^{\prime}v)=\omega_{\alpha 2}^{\prime}(q^{\prime})/\gamma - q_xv$, 
where $q_x^{\prime}=(q_x + \beta k)\gamma, \,\gamma=1/\sqrt{1-\beta^2},\,
\beta = v/c,\,k=\omega_{\alpha 1}(q)/c$.
Anomalous Doppler effect corresponds to the situation when $\omega_{\alpha 2}(q)<0$. 
In this case an excitation in the rest frame $K^{\prime}$ of body \textbf{2} 
corresponds to a gain of energy in the
\textit{lab} frame $K$. Thus, body \textbf{2} can radiate photons  while it is excited. Resonance 
arises when a gain of energy resulting from excitation of  body 2
with frequency $\omega_{gain}=-\omega_{\alpha 2}(\mathbf{q})$ 
will create excitation in the body \textbf{1} with 
frequency $\omega_{\alpha 1}(\mathbf{q})=-\omega_{\alpha 2}$ . 
Thus, the resonance condition has the form
\begin{equation}
\omega_{\alpha 1}(q)= q_xv- \omega_{\alpha 2}^{\prime}(q^{\prime})/\gamma.
\label{one}
\end{equation}
Eq. (\ref{one}) determines the condition for excitation of pair excitations. For transparent 
dielectrics such excitations can be photons. For this case Eq. (\ref{one}) determines occurrence of 
Vavilov-Cherenkov radiation. As result of such radiation a pair of photons (a photons with momentum 
$\mathbf{-q}$ in one body and a photon with momentum $\mathbf{q}$ in the other body) are created what give rise to the change of the momentum of each body  and 
friction which we denote as quantum friction due to the quantum origin of involved processes. 
For materials with losses also off-resonant processes  are possible  when an excitation is created 
in one body and resulting photon is absorbed by other body. 
The condition for such off-resonanant process in the non-relativistic case has the form
\begin{equation}
\omega_2(q)-q_xv<0
\label{condition0}
\end{equation}
Eq.(\ref{condition0}) was used by Landau for obtaining the critical velocity of a superfluid 
flowing past a wall \cite{Landau1941}.

For two transparent identical media $\omega_{\alpha 1}(q)=v_0q$ and 
$\omega_{\alpha 2}^{\prime}(q^{\prime})=v_0q^{\prime}$, where $v_0=c/n$, $n$ is 
the refractive index. In this case  the condition (\ref{one}) is reduced to the form
\begin{equation}
q_xv=\omega_{\alpha 1}(q)+\frac{\omega_{\alpha 2}^{\prime}(q^{\prime})}{\gamma}>v_0q_x\left(2-\frac{vv_0}{c^2}\right)
\label{condition1}
\end{equation}
or
\begin{equation}
v>v_c=\frac{2v_0}{1+(v_0/c)^2}=\frac{2nc}{n^2+1}.
\label{condition}
\end{equation}
The condition (\ref{condition}) was already obtained by Pendry \cite{PendryJMO1998} using 
an approximate relativistic theory. For $n\gg 1$ the condition (\ref{condition}) 
reduces to the non-relativistic result  $v>v_c=2c/n$ obtained in Ref. 
\cite{KardarPRA2013}. Thus, the condition for the validity of the non-relativistic
 theory is $v_0/c\ll 1$ or $n\gg 1$. However, for  transparent dielectrics, at 
the high frequencies typically involved,  $n\sim 1$ and a relativistic theory should be used.

Multiplying Eq.(\ref{one}) with the Plank constant and   with the photon emission rate per unit 
area, and taking into account that this rate is invariant  under the Lorentz transformation, we get
\begin{equation}
Fv=P_1 + \frac{P_2^{\prime}}{\gamma},
\label{three}
\end{equation}
where $F$ is the friction force, $P_1$ is the power of  photon emission energy, 
which is equal to the  power  of excitation energy in  the body \textbf{1} in the $K$-system, 
and $P_2^{\prime}$ is the power  of excitation energy in the body \textbf{2} in the $K^{\prime}$-system. For the lossy materials the energy absorbed by the bodies is converted into heat. 
In this case $P_1$ and $P_2^{\prime}$ are equal to the heat power in the corresponding reference frames.

\section{A fully relativistic theory for the Casimir friction between two plates sliding relative to each other}

According to a fully relativistic theory     \cite{VolokitinPRB2008}    the contributions  of the evanescent waves (which dominate at  large velocities and low temperatures)  to the friction force 
$F_{1x}$, and the radiation power  $P_1$ absorbed by plate  \textbf{1} in the    $K$ frame, are determined by formulas
\begin{equation}
\left(\begin{array}{c}
F_{1x}\\
P_1
\end{array} \right)
 = \int \frac{d^2q}{(2\pi)^2}
\int_0^{cq} \frac{d\omega}{2\pi} \left(\begin{array}{c}
\hbar q_x\\
\hbar \omega
\end{array} \right)\Gamma(\omega, q)\mathrm{sgn}(\omega^{\prime})[n_2(\omega^{\prime}) - n_1(\omega))]
\label{qvc1}
\end{equation}
where the positive quantity
\[
\Gamma_{12}(\omega, \mathbf{q})= \frac{4\mathrm{sgn}(\omega^{\prime})}{|\Delta|^2} [(q^2 - \beta kq_x)^2 -
\beta^2k_z^2q_y^2]
\{\mathrm{Im}R_{1p}[(q^2 - \beta kq_x)^2\mathrm{Im}R_{2p}^{\prime}|\Delta_{ss}|^2
\]
\begin{equation}
+ \beta^2k_z^2q_y^2
\mathrm{Im}R_{2s}^{\prime}|\Delta_{sp}|^2]
 +(p\leftrightarrow s)\}e^{-2 k_z d}
\end{equation}
can be identified as a spectrally resolved photon emission rate,
\[
\Delta = (q^2 - \beta kq_x)^2\Delta_{ss}\Delta_{pp} - \beta^2k_z^2q_y^2\Delta_{ps}\Delta_{sp},\,\,
\Delta_p=,
\]
\[
\Delta_{pp} = 1 - e^{-2k_zd}R_{1p}R_{2p}^{\prime},\, \Delta_{sp} = 
1 + e^{-2k_zd}R_{1s}R_{2p}^{\prime},
\]
$n_i(\omega)=[\exp(\hbar\omega/k_BT_i)-1]^{-1}$, $k_z=\sqrt{q^2-(\omega/c)^2}$, 
$R_{1 p(s)}$ is the reflection amplitude for surface
\textbf{1} in the  $K$ frame for   a $p(s)$ - polarized electromagnetic wave, 
$R_{2 p(s)}^{\prime} = R_{2 p(s)}(\omega^{\prime}, q^{\prime})$ is the reflection 
amplitude for surface
\textbf{2}  in the $K^{\prime}$ frame for   a $p(s)$ - polarized electromagnetic wave, 
$\omega^{\prime}=\gamma(\omega-q_xv)$, $q_x^{\prime}=\gamma (q_x-
\beta k)$,  $\Delta_{ps}=\Delta_{sp}(p\leftrightarrow s)$. 
The symbol
$(p\leftrightarrow s$) denotes the terms that are obtained from the preceding terms by 
permutation of indexes   $p$ and $s$. In the domains of the \textit{normal} 
Doppler effect  ($\omega - q_xv >0$) the last factor in the integrand in Eq. 
(\ref{qvc1}) can be written in the form
\[
\mathrm{sgn}(\omega^{\prime})[n_2(\omega^{\prime}) - n_1(\omega)] = n_2(\omega^{\prime})[1+n_1(\omega))] - n_1(\omega)[1+n_2(\omega^{\prime})].
\]
Thus, in this domain  the energy and momentum transfer are related as when the excitations are annihilated in one body and  created in other body. Such processes are only possible at  $T \ne 0$ K, i.e. they are associated with thermal radiation.  On the other hand, in the case of the \textit{anomalous} Doppler effect   ($\omega - q_xv <0$)
\[
\mathrm{sgn}(\omega^{\prime})[n_2(\omega^{\prime}) - n_1(\omega))] = 1 + n_2(|\omega^{\prime}|)+n_1(\omega) = [1 + n_2(|\omega^{\prime}|)][1 + n_1(\omega)] -n_2(|\omega^{\prime}|)n_1(\omega).
\]
In this case the excitations are created and annihilated  simultaneously in both bodies. 
Such processes are possible even at  $T = 0$ K and are associated with quantum friction.

If in Eq. (\ref{qvc1}) one
neglects  terms of  order $\beta^2$ then the contributions from waves
with $p$- and $s$- polarization will be separated. In this case
(\ref{qvc1}) is reduced to the formula obtained in Ref.  \cite{VolokitinJPCM1999}
\begin{equation}
\left(\begin{array}{c}
F_{1x}\\
P_1
\end{array} \right)
 = \frac 1 {2\pi ^3}\int d^2q
\int_0^{cq} d\omega \left(\begin{array}{c}
\hbar q_x\\
\hbar \omega
\end{array} \right)e^{-2 k_z d}
\left(\frac{\mathrm{Im}R_{1p}\mathrm{Im}R_{2p}^{\prime}}{|\Delta_{pp}|^2} + 
\frac{\mathrm{Im}R_{1s}\mathrm{Im}R_{2s}^{\prime}}{|\Delta_{ss}|^2}\right)
\left( n_2(\omega^{\prime})-n_1(\omega )\right),
\label{qvc2}
\end{equation}
Thus,  to  order  $\beta^2$ the mixing of the waves with 
the different polarization can be neglected what agree with the results obtained 
in Ref. \cite{VolokitinPRB2008}. At  $
T_1=T_2=0$ K the propagating waves do not contribute to the friction and the radiative 
heat transfer. However, the contribution from the evanescent waves does not vanish. 
Taking into account that  $n_1(\omega )=0$ at $
T_1=T_2=0$ K and $\omega >0$, 
\[
n_2(\omega^{\prime})=\left\{
\begin{array}{rl}
-1 & \text{for} \,\,0<\omega < q_xv \\
0& \text{for} \,\,\omega > q_xv
\end{array}\right.
\]
from Eq.  (\ref{qvc1}) we get the friction mediated by the evanescent waves at zero temperature  
(denoted  as quantum friction \cite{PendryJPCM1997})  and the radiative heat transfer
\begin{equation}
\left(\begin{array}{c}
F_{1x}\\
P_1
\end{array} \right)
 = \int_{-\infty}^{\infty} \frac{dq_y}{2\pi}\int_{0}^{\infty} \frac{dq_x}{2\pi}
\int_0^{q_xv} \frac{d\omega}{2\pi} \left(\begin{array}{c}
\hbar q_x\\
\hbar \omega
\end{array} \right)\Gamma_{12}(\omega, \mathbf{q}).
\label{qvc3}
\end{equation}
If in Eq. (\ref{qvc3}) one neglects by the terms of  order   $\beta^2$, then the contribution 
from the waves with
 $p$- and $s$- polarization will be separated. In this case  (\ref
{qvc3}) is reduced to the approximate (relativistic) formula used by Pendry in Ref. \cite{PendryJMO1998}
\begin{equation}
\left(
\begin{array}{c}
F_x\\
P_1
\end{array} \right)
=
-\frac \hbar {\pi ^3}\int_0^\infty dq_y \int_0^\infty dq_x
\int_0^{q_xv} d\omega \left(\begin{array}{c}
\hbar q_x\\
\hbar \omega
\end{array} \right)
\left(\frac{\mathrm{Im}R_{1p}\mathrm{Im}R_{2p}^{\prime}}{|D_{pp}|^2} + 
\frac{\mathrm{Im}R_{1s}\mathrm{Im}R_{2s}^{\prime}}{|D_{ss}|^2}\right)e^{-2 k_z d},
\label{approximate}
\end{equation}
In the non-relativistic and non-retarded limit, which can be formally obtained in the limit  
$c\rightarrow \infty$, Eq.(\ref{approximate}) is reduced to  the formula    obtained by 
Pendry in Ref. \cite{PendryJPCM1997}.

\section{Quantum friction between two transparent plates}

For the transparent dielectrics the reflection amplitudes are given by the Fresnel's formulas
\begin{equation}
R_p = \frac{in^2k_z-\sqrt{n^2(\omega/c)^2-q^2}}{in^2k_z+\sqrt{n^2(\omega/c)^2-q^2}},\,R_s = \frac{ik_z-\sqrt{n^2(\omega/c)^2-q^2}}{ik_z+\sqrt{n^2(\omega/c)^2-q^2}}.
\label{amplitude}
\end{equation}
In this case the friction force can be written in the form 
\begin{equation}
F_{1x}=\frac{\hbar v_0}{d^4}\tilde{g}\left(\frac{v}{v_0},\frac{v}{c}, n\right),
\label{rel}
\end{equation}
Where $\tilde{g}$ is a function of two dimentionless velocity ratios, and the refractive index $n$. 
In the non-relativistic limit 
($\beta^2\ll 1$), the dependence on vacuum velocity $c$ drops out and 
\begin{equation}
F^{nrel}_{1x}=\frac{\hbar v_0}{d^4}\left[g_s\left(\frac{v}{v_0}\right)+g_p\left(\frac{v}{v_0},n
\right)\right],
\label{nonrel}
\end{equation}
where the $s$-wave contribution $g_s$ depends only on the ratio of the velocity 
$v$ to the light speed in the medium $v_0$. This result was already 
noted in Ref. \cite{KardarPRA2013}. However, the $p$-wave contribution depends also on the 
refractive index $n$.

The imaginary part of the reflection amplitude $R_{1p(s)}$, given by Eq.(\ref{amplitude}),  
is only nonzero when $\omega > v_0q>v_oq_x$. Similarly, $\mathrm{Im}R_{2p(s)}$ is nonzero only when
$q_xv-\omega>v_0q^{\prime}/\gamma>v_0(q_x-\beta\omega/c)$. Both these conditions limit the range of integration to
\begin{equation}
v_0q_x<\omega<\frac{(v-v_0)q_x}{1-vv_0/c^2}.
\label{condition2}
\end{equation}
From this condition follows that the minimal velocity $v_c$,  at which  friction occurs,   
is determined by  Eq. (\ref{condition}). 

For  transparent dielectrics, where there are no resonances in the reflection amplitudes, in the frequency range where the quantum friction is non-vanishing 
$|R_{s(p)}|\leq 1$. Thus a good estimation of the friction force and 
the radiative heat transfer can be obtained by neglecting  the multiple-scattering of the electromagnetic waves by the dielectric surfaces, in the vacuum gap between them.  In this approximation 
$D_{pp}\approx D_{ss}\approx D_{sp} \approx D_{sp}\approx 1$,
\[
\Delta \approx (q^2 - \beta kq_x)^2 - \beta^2k_z^2q_y^2= \frac{(qq^{\prime})^2}{\gamma^2}, 
\]
\[
(q^2 - \beta kq_x)^2\mathrm{Im}R_{2p}^{\prime}|\Delta_{ss}|^2
+ \beta^2k_z^2q_y^2
\mathrm{Im}R_{2s}^{\prime}|\Delta_{sp}|^2\approx\frac{(qq^{\prime})^2}{\gamma^2} \mathrm{Im}R_{2p}^{\prime}+
\beta^2k_z^2q_y^2\mathrm{Im}(R_{2p}^{\prime}+R_{2s}^{\prime}),
\] 
\[
\Gamma_{12}=4\mathrm{sgn}(\omega^{\prime})
\left[\left(\mathrm{Im}R_{1p}\mathrm{Im}R_{2p}^{\prime}+\mathrm{Im}R_{1s}\mathrm{Im}R_{2s}^{\prime}\right)\left(1+\gamma^2\beta^2\frac {k_z^2q_y^2}{q^2q^{\prime 2}}\right)\right.
\]
\begin{equation}
\left.+\gamma^2\beta^2\frac {k_z^2q_y^2}{q^2q^{\prime 2}}\left(\mathrm{Im}R_{1p}\mathrm{Im}R_{2s}^{\prime}+\mathrm{Im}R_{1s}\mathrm{Im}R_{2p}^{\prime}\right)\right].
\label{gamma}
\end{equation}
Thus the relativistic effects produce not only a mixing of the waves with the different polarization, but also modify the contributions 
from the different polarizations. These effects were not taken into account in the approximate  relativistic theory used by Pendry \cite{PendryJMO1998}.

Close to the threshold velocity $v\approx v_c$, when $\xi_{min}\approx \xi_{max}$ and 
\[
y_{max}=\frac{n^2+1}{n\sqrt{n^2-1}}\sqrt{\frac{v-v_c}{v_0}}\ll 1,
\]
the integration over $q_y$ in Eq. (\ref{qvc3}) is restricted by the range $0<|q_y|<y_{max}q_x\ll q_x$. In this case, 
to lowest order in $y_{max}$, the mixing of the waves with different polarization can be neglected and the 
friction force is determined by formula (see Appendix \ref{A})
\begin{equation}
F_{1x}\approx\frac{\hbar v_0}{d^4}\left[\tilde{g_s}\left(\frac{v}{v_0},n\right)+\tilde{g_p}\left(\frac{v}{v_0},n\right)\right],
\label{qfc5}
\end{equation}
and the radiative heat transfer $P_1=v_0F_{1x}$, where
\[
\tilde{g_s}\left(\frac{v}{v_0},n\right)=\frac{\zeta(3)}{5\pi^2}\frac{n(n^2+1)^5}{(n^2-1)^5\sqrt{n^2-1}}\left(\frac{v-v_c}{v_0}\right)^{5/2},
\]
and $\tilde{g}_p=\tilde{g}_s/n^4$. In the non-relativistic limit ($n\gg 1$)
\[
g_s\left(\frac{v}{v_0}\right)=\frac{\zeta(3)}{5\pi^2}\left(\frac{v-v_c}{v_0}\right)^{5/2},
\]
and $g_p(v/v_0,n)=g_s(v/v_0)/n^4$.

\begin{figure}
\includegraphics[width=0.5\textwidth]{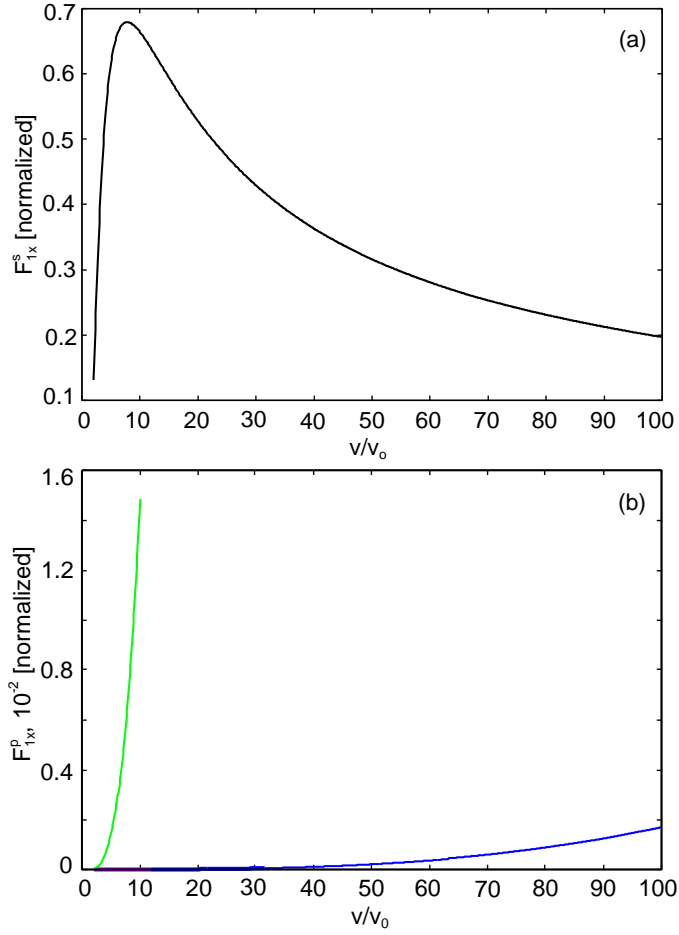}
\caption{\label{Fig.2.} The dependence of friction force between 
two  transparent dielectric plates  on the relative sliding velocity. 
Normalization factor for the forces $\hbar v_0/\pi^3d^4,\,v_0=c/n$. (a) and (b): results of a 
non-relativistic 
theory 
for the contributions from $s$- and $p$-polarised electromagnetic waves,  respectively. 
The $s$-wave  contribution in a non-relativistic theory depends only 
on the ratio $v/v_0$.  The $p$-wave contributions are shown for $n=10$ (green line) and $n=100$ (blue line). 
 }.
\end{figure}

\begin{figure}
\includegraphics[width=0.5\textwidth]{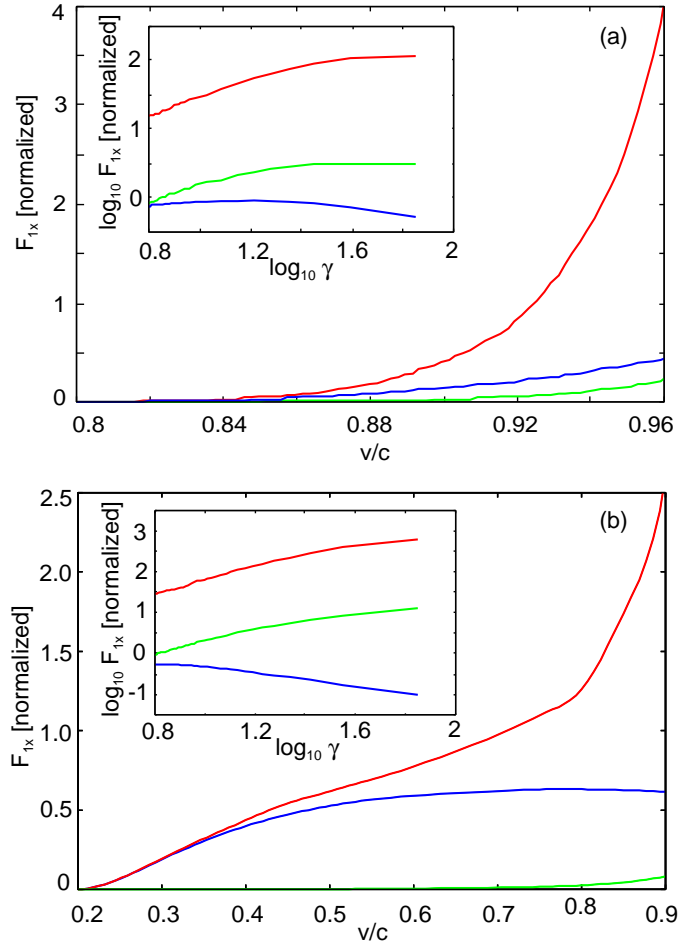}
\caption{\label{Fig.3.} The same as on Fig. \ref{Fig.2.} but for a fully relativistic theory.
Figures (a) and (b)  show 
the results of a fully relativistic theory (red line) for $n=2$  and $n=10$, respectively.  
The blue and green  lines show the separate contributions from the $s$- and 
$p$-polarized electromagnetic waves, respectively, obtained using the approximate formula (\ref{approximate}). 
Insets show the friction forces in the ultrarelativistic case ($1-\beta\ll 1$) }.
\end{figure}

Close to the light velocity ($\gamma \gg 1$) (see Appendix B)
the $s$-wave contributions to the friction force, given by  the approximate formula (\ref{approximate}), is finite at $v\rightarrow c$ 
\begin{equation}
F_{xs}^{approx}\approx  \frac {3\hbar v}{4\pi ^2d^4}\frac{\sqrt{2}}{n^2-1}\mathrm{ln}(n+\sqrt{n^2-1}),
\end{equation}
and  diverges as $\sim \gamma$ in a fully relativistic theory given by Eq. (\ref{qvcapprox}),
\begin{equation}
F_{xs}\approx \frac {3\hbar v}{4\pi ^2d^4}\frac{\sqrt{n-1}}{2(n+1)^{3/2}}\gamma
\end{equation}
Other contributions can be estimated in a similar way.

Figs. \ref{Fig.2.} and \ref{Fig.3.} show  the dependence of friction force between two transparent dielectric  plates  
on the relative sliding velocity in a non-relativistic theory 
Figs.\ref{Fig.2.}(a-b), and  a fully relativistic theory for $n=2$ Fig. \ref{Fig.3.}(a) and $n=10$ Fig. \ref{Fig.3.}(b). In a 
non-relativistic theory the contributions 
to the friction force from $s$-and $p$-polarised waves  are separated. The threshold velocity
$v_c$ for appearance of the Vavilov-Cherenkov radiation in the non-relativistic theory is equal 
to 2$v_0$. The friction in this theory is dominated by the $s$-wave contribution which
depends only on the velocity ratio $v/v_0$. 
In a fully  relativistic theory   friction and radiation  only exist for  $v >v_c=2nc/(n^2+1)$, which is equal to 0.8$c$ 
for $n=2$ Fig. \ref{Fig.3.}(a) 
and 0.2$c$ for  $n=10$ Fig. \ref{Fig.3.}(b). Figs. 3(a-b) also show results of an approximate relativistic theory 
for the contributions to the friction force from
$s$-polarised waves (blue line) and $p$-polarised waves (green line) given by Eq. (\ref{approximate}). In such an approximate theory the reflection amplitude 
from the moving surface is approximated by the reflection amplitude in the co-moving reference frame at the frequencies  and wave-vectors determined 
by the Lorenz transformation. The polarization mixing is not taken into account in this approximated theory,  
where the coupling between  waves with different polarization are neglected.   
Close to the threshold velocity the mixing of the waves with different polarization is 
unimportant and the friction is dominated   by the contribution 
from the $s$--polarized electromagnetic waves, which can be accurately described using an approximate theory. However, in the ultrarelativistic case 
($\gamma \gg 1$) both contributions from the different polarizations are strongly enhanced in  the comparison with the approximate theory, and a new contribution 
occurs connected with the polarization mixing.

\section{Conclusion}
Two   transparent dielectric plates  during relative sliding emits   quantum Valivov-Cherenkov radiation when the sliding velocity exceeds 
a threshold velocity $v_c=2nc/(n^2+1)$. This radiation is responsible for quantum friction which we have studied using a fully relativistic theory. Close to the threshold velocity the 
friction force 
$\sim (v-v_c)^{5/2}$ and is dominated by the contribution from the $s$-polarized electromagnetic waves. However, close to the light velocity the contributions from both 
polarization are strongly enhanced and a new contribution occurs connected with  the mixing of the waves with the 
different polarization. As  was shown by Pendry \cite{PendryJMO1998}, surface roughness also can strongly enhance the Vavilov-Cherenkov radiation. In the relativistic case this 
problem requires further investigations.

\section*{Acknowledgement}
The study was supported by the Ministry of education and science of Russia under Competitiveness Enhancement Program of SSAU for 2013-2020 years, the Russian Foundation for 
Basic Research (Grant No. 14-02-00384-a) and COST Action MP1303 “Understanding and Controlling Nano and Mesoscale Friction.”

\appendix

\section{Close to the threshold velocity \label{A}}

Close to the threshold ($(v-v_c)/v_0\ll 1$) the range in $\omega$ becomes narrow. For small $\omega-q_xv$: 
\[
k^2_{zn}=n^2\left(\frac{\omega}{c}\right)^2-q^2=\left[\frac{(\omega-q_xv_0)(\omega+q_xv_0)}{v_0^2}-q_y^2\right]
\]
\begin{equation}
\approx \left(\frac{\omega-q_xv_0}{v_0}\right)2q_x-q_y^2,
\label{a1}
\end{equation}

\[
k_{nz}^{\prime 2}=\left(\frac{\omega^{\prime}}{v_0}\right)^2-q^{\prime 2}
\]
\[
=-\gamma^2 \frac{[q_x(v-v_0)-\omega(1-vv_0/c^2)][q_x(v+v_0)-\omega(1+vv_0/c^2]}{v^2_0}-q_y^2
\]
\begin{equation}
=\frac{1}{v_0}\left[\frac{(n^2+1)^2}{n^2(n^2-1)}(v-v_c)q_x-(\omega-q_xv_0)\right]2q_x-q_y^2.
\label{a2}
\end{equation}
From Eqs. (\ref{a1}) and (\ref{a2}) follow the ranges in $\omega$ and $q_y$: $\omega_{-}<\omega<\omega_{+}$ where
\[
\omega_{-}=q_xv_0+\frac{ v_0q_y^2}{2q_x},\,\ \omega_+=q_xv_0+\frac{(n^2+1)^2}{n^2(n^2-1)}(v-v_c)q_x-\frac{ v_0q_y^2}{2q_x},
\]
and $0\leq q_y\leq q_{max}$, where
\[
q_{max}^2=\frac{(n^2+1)^2}{n^2(n^2-1)}\frac{v-v_c}{v_0}q_x^2 \ll q_x^2.
\]
After  the changing of the variables $q_y\rightarrow q_xy_{max}y$, 
\[
\omega \rightarrow q_xv_0\left(1+\frac{y_{max}^2}{2}+zy_{max}^2\frac{1-y^2}{2}\right),
\]
where 
\[
y_{max}^2=\frac{(n^2+1)^2}{n^2(n^2-1)}\frac{v-v_c}{v_0}, 
\]
the imaginary parts for the reflection amplitudes can be written in the form
\begin{equation}
\mathrm{Im}R_s=\frac{2k_zk_{zn}}{k_z^2+k_{zn}^2}\approx \frac{2k_{zn}}{k_z}\approx \frac{2n}{q_x\sqrt{n^2-1}}\sqrt{\left(\frac{\omega-q_xv_0}{v_0}\right)2q_x-q_y^2}
=\frac{2ny_{max}}{\sqrt{n^2-1}}\sqrt{1-y^2}\sqrt{1+z} \sim \sqrt{\frac{v-v_c}{v_0}},
\label{a3}
\end{equation}
\[
\mathrm{Im}R_s^{\prime}=\frac{2k_zk_{zn}^{\prime}}{k_z^2+k_{zn}^{\prime 2}}\approx\frac{2k_{zn}^{\prime}}{k_z}\approx \frac{2n}{q_x\sqrt{n^2-1}}\sqrt{\frac{1}{v_0}\left[\frac{(n^2+1)^2}{n^2-1}(v-v_c)q_x-(\omega-q_xv_0)\right]2q_x-q_y^2}
\]
\begin{equation}
=\frac{2ny_{max}}{\sqrt{n^2-1}}\sqrt{1-y^2}\sqrt{1-z} \sim \sqrt{\frac{v-v_c}{v_0}},
\label{a4}
\end{equation}
$\mathrm{Im}R_p\approx \mathrm{Im}R_s/n^2$, $\mathrm{Im}R_p^{\prime}\approx \mathrm{Im}R_s^{\prime}/n^2$. Because the integrand in Eq. (\ref{qvc3}) is proportional to 
the product of the imaginary parts of the reflection amplitudes, 
which are  of the order $(v-v_c)/v_0$, to lowest order in $(v-v_c)/v_0$, all other terms in the integrand should be taken at $v=v_c$. In this approximation 
the mixing of the waves with the different polarization can be neglected because they are of  
 order $q_y^2\sim (v-v_c)/v_0$ and the reflection amplitudes $R_s=R_p=1$ and  
the integral for the contribution to the friction force from $s$-polarized waves  is reduced to
\begin{equation}
F_{1x}^s\approx \frac{\hbar v_0}{\pi^3}\int_0^{\infty}dq_xq_x^3\frac{e^{-2q_xd\sqrt{n^2-1}/n}}{(1- e^{-2q_xd\sqrt{n^2-1}/n})^2}\left(\frac{2n}{\sqrt{n^2-1}}\right)^2
y_{max}^5\int_0^1dy(1-y^2)^2 \int_{-1}^{1}dz\sqrt{1-z^2}
\end{equation}
which produces Eq. (\ref{qfc5}).

\section{Close to the light velocity: $v \rightarrow c$ \label{B}}

Introducing new variables $\omega = q_xv\xi$ and $q_y=q_xy$ the integration in Eq. (\ref{gamma}) over $q_x$ can be performed analytically giving 
\[
\left(
\begin{array}{c}
F_x\\
P_1
\end{array} \right)
=
-\frac {3\hbar v}{8\pi ^3d^4}\int_{\xi_{min}}^{\xi_{max}} d\xi 
\int_0^{y_{max}} dy \left(\begin{array}{c}
1\\
 v \xi
\end{array} \right)
\frac{1}{\kappa_z^4}\left[\left(\mathrm{Im}R_{1p}\mathrm{Im}R_{2p}^{\prime} + 
\mathrm{Im}R_{1s}\mathrm{Im}R_{2s}^{\prime}\right)\left(1+\gamma^2\beta^2\frac {\kappa_z^2y^2}{w^2w^{\prime 2}}\right)\right.
\label{approximate1}
\]
\begin{equation}
\left.+\gamma^2\beta^2\frac {\kappa_z^2y^2}{w^2w^{\prime 2}}\left(\mathrm{Im}R_{1p}\mathrm{Im}R_{2s}^{\prime} + 
\mathrm{Im}R_{1s}\mathrm{Im}R_{2p}^{\prime}\right)\right]
\label{qvcapprox}
\end{equation}
where: $
\kappa_z^2= 1-\beta^2\xi^2+y^2
$, $w^2=1+y^2$, $w^{\prime 2}=\gamma^2(1-\beta^2\xi)^2+y^2$, 
\[
\xi_{min}=\frac{1}{n\beta}, \,\,\xi_{max}=1-\frac{1}{\gamma^2\beta (n-\beta)}=\frac{n\beta -1}{\beta(n-\beta)}.
\]

In new variables $k_z=q_x\kappa_{z}$, 
\[
k_{nz}^2=(n^2-1)\left(\frac{\omega}{c}\right)^2-k_z^2= q_x^2(y_1^2-y^2),
\]
\[
k_{nz}^{\prime 2}=(n^2-1)\left(\frac{\omega^{\prime}}{c}\right)^2-k_z^2= q_x^2(y_0^2-y^2),
\]
where $y_1=n^2\beta^2\xi^2-1$,
\[
y_{0}^2=\gamma^2[n^2\beta^2(1-\xi)^2-(1-\beta^2\xi)^2]
\]
\[
\mathrm{Im}R_s=\frac{2k_zk_{nz}}{k_z^2+k_{nz}^2}=\frac{2\sqrt{1-\beta^2\xi^2+y^2}\sqrt{y_1^2-y^2}}
{(n^2-1)\beta^2\xi^2},
\]\[
\mathrm{Im}R_s^{\prime}=\frac{2k_zk_{nz}^{\prime}}{k_z^2+k_{nz}^{\prime 2}}=\frac{2\sqrt{1-\beta^2\xi^2+y^2}\sqrt{y_0^2-y^2}}
{(n^2-1)\gamma^2\beta^2(1-\xi)^2},
\]
\[
\mathrm{Im}R_p=\frac{2n^2k_zk_{nz}}{n^4k_z^2+k_{nz}^2}=\frac{2n^2\sqrt{1-\beta^2\xi^2+y^2}\sqrt{n^2\beta^2\xi^2-1-y^2}}
{(n^2-1)\beta^2\xi^2+(n^4-1)(1-\beta^2\xi^2+y^2)},
\]
\[
\mathrm{Im}R_p^{\prime}=\frac{2n^2k_zk_{nz}}{n^4k_z^2+k_{nz}^2}=\frac{2n^2\sqrt{1-\beta^2\xi^2+y^2}\sqrt{y_0^2-y^2}}
{y_{0}^2 + (n^4-1)(1-\beta^2\xi^2+y^2) +1-\beta^2\xi^2},
\]
\[
y_{max}=\left\{
\begin{array}{rl}
y_0 & \text{for} \,\,\xi_c<\xi < \xi_{max} \\
y_1& \text{for} \,\,\xi_{min}<\xi<\xi_c
\end{array}\right.,
\]
where $\xi_c=\gamma/(1+\gamma)\approx 1-1/\gamma$. At $\gamma \gg 1$ the main contribution during integration gives the region $\xi_c<\xi< \xi_{max}$. The $s$-wave contribution 
is given by
\[
\left(
\begin{array}{c}
F_{xs}\\
P_{1s}
\end{array} \right)
\approx
\frac {3\hbar v}{8\pi ^3d^4}\int_{\xi_c}^{\xi_{max}} d\xi 
\int_0^{1} dy \left(\begin{array}{c}
1\\
 v 
\end{array} \right)
\frac{4y_0^2\sqrt{1-y^2}}{(n^2-1)^{3/2}\gamma^2(1-\xi)^2}
\]
\[
\times\left\{\frac{1}{1-\beta^2\xi^2+y_0^2y^2}+\gamma^2\frac{y_0^2y^2}{(1+y_0^2y^2)[\gamma^2(1-\beta^2\xi)^2+y_0^2y^2]}\right\}
\]
\[
\approx \frac {3\hbar v}{4\pi ^2d^4}\left(\begin{array}{c}
1\\
 v 
\end{array} \right)
\left\{\left[\frac{\sqrt{2}}{n^2-1}\mathrm{ln}(n+\sqrt{n^2-1})- \frac{1}{\sqrt{n^2-1}(n+1)}-\frac{2}{(n^2-1)\sqrt{\gamma}}
+\frac{1}{(n^2-1)^{3/2}\gamma}\right]\right.
\]
\begin{equation}
\left.+\left[\frac{\sqrt{n-1}}{2(n+1)^{3/2}}\gamma-\frac{1}{\sqrt{n^2-1}(n+1)}\textrm{ln}(n-1)\gamma - \frac{1}{2(n^2-1)^{3/2}\gamma} \right]\right\}.
\end{equation}
Other contributions can be estimated in a similar way.

\end{document}